\documentclass[twocolumn,aps,showpacs,superscriptaddress,floatfix,
               preprintnumbers,amsmath,amssymb,nofootinbib]{revtex4}

\usepackage{graphicx}
\usepackage{dcolumn}
\usepackage{bm}

\newcommand{\bear}{\begin{eqnarray}}    
\newcommand{\eear}{\end{eqnarray}}      
\newcommand{\beqstar}{\begin{eqnarray*}}        
\newcommand{\eeqstar}{\end{eqnarray*}}

\def\eslt{E_T^{miss}}
\def\to{\rightarrow}

\def\bi{\begin{itemize}}
\def\ei{\end{itemize}}
\def\te{\tilde e}

\def\tu{\tilde u}
\def\sps1ap{SPS1a$^\prime$}
\def\c1p{C1$^\prime$}

\def\ta{\tilde a}
\def\tb{\tilde b}

\def\tst{\tilde t}
\def\ttau{\tilde \tau}

\def\tg{\tilde g}
\def\tnu{\tilde\nu}
\def\tell{\tilde\ell}
\def\tq{\tilde q}
\def\tw{\widetilde W}
\def\tz{\widetilde Z}
\def\alt{\stackrel{<}{\sim}}
\def\agt{\stackrel{>}{\sim}}
\def\be{\begin{equation}}  
\def\ee{\end{equation}}  
\def\bea{\begin{eqnarray}}  
\def\eea{\end{eqnarray}}  
\def\beas{\begin{eqnarray*}}  
\def\eeas{\end{eqnarray*}}  
\newcommand\plb[3]{{Phys.\ Lett.\ }{\bf B #1}, #2 (#3)}
\newcommand\jhep[3]{{J. High Energy Phys.\ }{\bf #1}, #2 (#3)}

\newcommand\npb[3]{{\it Nucl.\ Phys.\ }{\bf B #1} (#2) #3}

\newcommand{\hepph}[1]{hep-ph/#1}

\newcommand\ppnp[3]{{\it Prog.\ Part.\ Nucl.\ Phys.}{\bf  #1} (#2) #3}

\begin{document}


\title{Radiative natural supersymmetry\\ 
with mixed axion/higgsino cold dark matter}

\author{Howard Baer}
\affiliation{Dept. of Physics and Astronomy,
University of Oklahoma, Norman, OK 73019, USA}

\date{January 16, 2009}

\begin{abstract}
Models of natural supersymmetry seek to solve the little hierarchy problem by positing a
spectrum of light higgsinos $\alt 200$ GeV and light top squarks
$\alt 500$ GeV along with very heavy squarks and TeV-scale gluinos. Such models have low
electroweak finetuning and are safe from LHC searches. However, in the context of the MSSM,
they predict too low a value of $m_h$ and the relic density of thermally produced higgsino-like WIMPs 
falls well below dark matter (DM) measurements. 
Allowing for high scale soft SUSY breaking Higgs mass $m_{H_u}> m_0$ 
leads to natural {\it cancellations} during RG running, and to radiatively induced {\it low finetuning at the electroweak scale}.
This model of {\it radiative natural SUSY} (RNS), with large mixing in the top squark sector, allows
for finetuning at the 5-10\% level with TeV-scale top squarks and a 125 GeV light Higgs scalar $h$.
If the strong CP problem is solved via the PQ mechanism, then we expect an axion-higgsino
admixture of dark matter, where either or both the DM particles might be directly detected.
\end{abstract}

\maketitle

\section{Introduction}

The recent fabulous discovery by Atlas and CMS of a Higgs-like resonance at 125 GeV\cite{atlas_h,cms_h} 
adds credence to supersymmetric models (SUSY) of particle physics in that the mass value 
falls squarely within the narrow predicted MSSM window: $m_h\sim 115-135$ GeV\cite{mhiggs}. 
At the same time, a lack of a SUSY signal at LHC7 and LHC8 implies squarks and gluinos
beyond the 1 TeV range\cite{atlas_susy,cms_susy}, exacerbating the {\it little hierarchy problem} (LHP): 
\bi
\item how do multi-TeV values of SUSY model parameters conspire to yield a $Z$-boson mass of just 91.2 GeV? 
\ei
Models of {\it natural supersymmetry}\cite{kn} address the LHP by positing a spectrum of light higgsinos 
$\alt 200$ GeV and light top squarks $\alt 500$ GeV along with very heavy first/second generation 
squarks and TeV-scale gluinos\cite{ah,others,bbht}. 
Such a spectrum allows for low electroweak finetuning while at the same time keeping sparticles
safely beyond LHC search limits.
In these models, the radiative corrections to $m_h$, which increase with $m_{\tst_i}^2$, 
are somewhat suppressed and have great difficulty in generating a
light SUSY Higgs scalar with mass $m_h\sim 125$ GeV\cite{h125}. 
Thus, we are faced with a new conundrum: how do we
reconcile low electroweak finetuning with such a large value of $m_h$\cite{hpr}?
In addition, the light higgsino-like WIMP particles predicted by models of natural SUSY 
lead to a thermally-generated relic density which is typically a factor
10-15 below\cite{bbh,bbht} the WMAP measured value of $\Omega_{CDM}h^2\simeq 0.11$.

One solution to the finetuning/Higgs problem is to add extra matter to the theory, 
thus moving beyond the MSSM\cite{hpr}. For example, adding 
an extra singlet as in the NMSSM adds further quartic terms to the higgs potential thus allowing for increased
values of $m_h$\cite{nmssm}. One may also add extra vector-like matter to increase $m_h$ while maintaining 
light top squarks\cite{vecmatter}. In the former case of the NMSSM, adding extra gauge singlets may lead to 
re-introduction of destabilizing divergences into the theory\cite{bpr}. In the latter case, one might wonder
about the ad-hoc introduction of extra weak scale matter multiplets and how they might have avoided detection
A third possibility, which is presented below, is to re-examine EWFT and to 
ascertain if there do exist sparticle spectra {\it within the MSSM} that lead to $m_h\sim 125$ GeV 
while maintaining modest levels of electroweak finetuning. 

\section{Electroweak finetuning}

One way to evaluate EWFT in SUSY models is to examine the minimization condition on the
Higgs sector scalar potential which determines the $Z$ boson mass. (Equivalently, one may examine
the mass formula for $m_h$ and draw similar conclusions.) 
One obtains the well-known tree-level expression
\be 
\frac{m_Z^2}{2} =
\frac{m_{H_d}^2-m_{H_u}^2\tan^2\beta}{\tan^2\beta -1} -\mu^2 \; .
\label{eq:treemin}
\ee 
To obtain a {\it natural} value of $M_Z$ on the left-hand-side, one would
like each term $C_i$ (with $i=H_d,\ H_u$ and $\mu$) on the right-hand-side to be of 
order $m_Z^2/2$. This leads to a finetuning parameter definition
\be
\Delta\equiv max_i \left(C_i\right)/(m_Z^2/2)
\ee
where $C_{H_u}=|-m_{H_u}^2\tan^2\beta /(\tan^2\beta -1)|/$, 
$C_{H_d}=|m_{H_d}^2/(\tan^2\beta -1)|/$ and $C_\mu =|-\mu^2|$ .
Since  $C_{H_d}$ is suppressed by $\tan^2\beta -1$, for even moderate $\tan\beta$
values this expression reduces approximately to
\be
\frac{m_Z^2}{2} \simeq -m_{H_u}^2-\mu^2 .
\label{eq:approx}
\ee
The question then arises: what is the model, what are the input parameters, and how do 
we interpret Eq's \ref{eq:treemin} and \ref{eq:approx}? 

Suppose we have a model with input parameters defined at some high scale $\Lambda\gg m_{SUSY}$, 
where $m_{SUSY}$ is the SUSY breaking scale $\sim 1$ TeV. Then
\be
m_{H_u}^2(m_{SUSY})=m_{H_u}^2(\Lambda )+\delta m_{H_u}^2
\ee
 where 
\be
\delta m_{H_u}^2\simeq -\frac{3f_t^2}{8\pi^2}\left(m_{Q_3}^2+m_{U_3}^2+A_t^2 \right)
\ln\left(\frac{\Lambda}{m_{SUSY}}\right) .
\ee
The usual lore is that in a model defined at energy scale $\Lambda$, then both
$m_{H_u}^2(\Lambda )$ and $\delta m_{H_u}^2$ must be of order $m_Z^2/2$ in order to avoid
finetuning. In fact, requiring $\delta m_{H_u}^2\alt m_Z^2/2$ has been used to argue
for a sparticle mass spectra of natural SUSY. Taking $\Delta =10$ corresponds to
\begin{itemize}
\item $|\mu | \alt 200\ {\rm GeV}$,
\item $m_{\tst_i},\ m_{\tb_1}\alt 500\ {\rm GeV}$,
\item $m_{\tg}\alt 1.5\ {\rm TeV}$.
\end{itemize}  
Since first/second generation squarks and sleptons hardly enter into Eq.~\ref{eq:treemin}, 
these can be much heavier: beyond LHC reach and also possibly providing a 
(partial) decoupling solution to the SUSY flavor and CP problems:
\begin{itemize}
\item $m_{\tq,\tell}\sim 10-50$ TeV.
\end{itemize}
The natural SUSY solution reconciles lack of a SUSY signal at LHC with allowing for
electroweak naturalness. It also predicts that the $\tst_{1,2}$ and $\tb_1$ may soon
be accessible to LHC searches. New limits from direct top and bottom squark pair production searches,
interpreted within the context of simplified models, are biting into the NS parameter space\cite{lhc_stop}. 
Of course, if $m_{\tst_{1,2}},\ m_{\tb_1}\simeq m_{\tz_1}$, then the visible decay products from
stop and sbottom production will be soft and difficult to see at LHC.
A more worrisome problem is that, with such light top squarks, the radiative corrections
to $m_h$ are not large enough to yield $m_h\simeq 125$ GeV. This problem has been used to argue
that additional multiplets beyond those of the MSSM must be present in order to raise up
$m_h$ while maintaining very light third generation squarks\cite{hpr}.
A third issue is that the relic abundance of higgsino-like WIMPs, calculated in the standard
MSSM-only cosmology, is typically a factor 10-15 below measured values.
These issues have led some people to grow increasingly skeptical of weak scale SUSY, 
even as occurs in the natural SUSY incarnation.

One resolution to the above finetuning problem is to merely invoke a SUSY particle spectrum at the 
weak scale, as in the pMSSM model. Here, $\Lambda\sim m_{SUSY}$ so $\delta m_{H_u}^2\sim 0$ and we
may select parameters $m_{H_u}^2\sim \mu^2\sim m_Z^2$. While a logical possibility, this
solution avoids the many attractive features of a model which is valid up to a high scale such as
$\Lambda\sim m_{GUT}$, with gauge coupling unification and radiative electroweak symmetry
breaking driven by a large top quark mass.

Another resolution is to impose Eq. \ref{eq:treemin} as a condition on high scale models, but using only
weak scale parameters. In this case, we will differentiate the finetuning measure as $\Delta_{EW}$, 
while the finetuning measure calculated using high scale input parameters we will refer to as
$\Delta_{HS}$. The weaker condition of allowing only for $\Delta_{EW}\alt 20$ allows for 
possible {\it cancellations} in $m_{H_u}^2(m_{SUSY})$. This is precisely what happens in what is known
as the hyperbolic branch or focus point region of mSUGRA\cite{hb_fp}: 
$m_{H_u}^2(\Lambda)\simeq -\delta m_{H_u}^2 \sim m_Z^2$ and consequently a value of $\mu^2\sim m_Z^2$ 
is chosen to enforce the measured value of $m_Z$ from Eq. \ref{eq:treemin}.\footnote{This may also occur in 
other varied models such as mixed moduli-AMSB\cite{nilles}.}
The HB/FP region of mSUGRA occurs for small values of trilinear soft parameter $A_0$. 
Small $A_0$ leads to small $A_t$ at the weak scale, which leads to $m_h\sim 115-120$ GeV, well below
the Atlas/CMS measured value of $m_h\simeq 125$ GeV. Scans over parameter space show the HB/FP 
region is nearly excluded if one requires both low $|\mu |\sim m_Z$ and $m_h\sim 123-127$ GeV\cite{bbm2,sug}.

The cancellation mechanism can also be seen from an approximate analytic solution 
of the EWSB minimation by Kane {\it et al.}\cite{kane}:
\bea
m_Z^2\simeq &=&-1.8\mu^2 +5.9 M_3^2- 0.4M_2^2-1.2m_{H_u}^2+0.9 m_{Q_3}^2\nonumber \\
& & +0.7 m_{U_3}^2 -0.6A_tM_3+0.4 M_2M_3+\cdots
\eea
(which adopts $\tan\beta =5$ although similar expressions may be gained for other $\tan\beta$ values). 
All parameters on the RHS are GUT scale parameters. We see one solution for
obtaining $m_Z$ on the left-hand side is to have all GUT scale parameters of order $m_Z$ (this is now excluded by
recent LHC limits). The other possibility-- 
if some terms are large (like $M_3\agt 0.4$ TeV in accord with recent LHC limits)-- is to 
have large cancellations. The simplest possibility-- using $M_3\agt 0.4$ TeV-- is then to raise up $m_{H_u}^2$
beyond $m_0$  such that there is a large cancellation. This possibility is allowed in the non-universal
Higgs models\cite{nuhm2,nuhm1}.

In mSUGRA, the condition that $m_{H_u}^2=m_{H_d}^2=m_{\tq}^2=m_{\tell}^2\equiv m_0^2$ at the 
high scale is anyways hard to accept. One might expect that all matter scalars
in each generation are nearly degenerate since the known matter in each generation fills out complete
16-dimensional representations of $SO(10)$. However, the distinguishing feature of the Higgs multiplets
is that they would live in 10-dimensional representations, and we then would {\it not} expect
$m_{10}=m_{16}$ at $m_{GUT}$. 
A more likely choice would be to move to the non-universal Higgs model, which comes in several 
varieties. It was shown in Ref. \cite{nuhm1} that by adopting a simple one-parameter 
extension of mSUGRA-- known as the one extra parameter non-universal Higgs model (NUHM1),
with parameter space
\be
m_{H_u}^2=m_{H_d}^2\equiv m_\phi^2,\ m_0,\ m_{1/2},\ A_0,\ \tan\beta ,\ sign(\mu)
\ee
-- then for any spectra one may raise $m_\phi$ up beyond $m_0$ until at some point $m_{H_u}^2(m_{SUSY})$
drops in magnitude to $\sim m_Z^2$. The EWSB minimization condition then also forces $|\mu |\sim m_Z$. 
The worst of the EWFT is eliminated due to a large cancellation between $m_{H_u}^2(\Lambda =m_{GUT})$ and
$\delta m_{H_u}^2$ leading to low $\Delta_{EW}$ and a model which enjoys electroweak naturalness.

The cancellation obviously can also be implemented in the 2-extra-parameter model NUHM2 where both
$m_{H_u}^2(m_{GUT})$ and $m_{H_d}^2(m_{GUT})$ may be taken as free parameters (as in an $SU(5)$ SUSY GUT)
or-- using the EWSB minimization conditions-- these may be traded for weak scale values of
$\mu$ and $m_A$ as alternative inputs\cite{nuhm2}. 
A third possibility that will allow for an improved 
decoupling solution to the SUSY flavor and CP problems would occur if we allow
for split generations $m_0(3)$ and $m_0(1)\simeq m_0(2)\equiv m_0(1,2)$ (SGNUHM). The latter 
condition need not require exact degeneracy, since with $m_0(1)\sim m_0(2)\sim 10-20$ TeV
we obtain only a {\it partial} decoupling solution to the flavor problem.
Taking $m_0(1,2)\sim 10-20$ TeV solves the SUSY CP problem\cite{susyflcp}.

\section{Radiative natural SUSY}

Motivated by the possibility of cancellations occuring in $m_{H_u}^2(m_{SUSY})$, we go back to the 
EWSB minimization condition and augment it with radiative corrections $\Sigma_u^u$ and $\Sigma_d^d$ since
if $m_{H_u}^2$ and $\mu^2$ are suppressed, then these may dominate:
\be 
\frac{m_Z^2}{2} =
\frac{(m_{H_d}^2+\Sigma_d^d)-(m_{H_u}^2+\Sigma_u^u)\tan^2\beta}{\tan^2\beta -1} -\mu^2 \; .
\label{eq:loopmin}
\ee 
Here the $\Sigma_u^u$ and $\Sigma_d^d$ terms arise from derivatives
of the {\it radiatively corrected} Higgs potential evaluated at the potential minimum.
At the one-loop level, $\Sigma_u^u$ contains the
contributions\cite{an} $\Sigma_u^u(\tst_{1,2})$, $\Sigma_u^u(\tb_{1,2})$,
$\Sigma_u^u(\ttau_{1,2})$, $\Sigma_u^u(\tw_{1,2})$,
$\Sigma_u^u(\tz_{1-4})$, $\Sigma_u^u(h,H)$, $\Sigma_u^u(H^\pm)$,
$\Sigma_u^u(W^\pm)$, $\Sigma_u^u(Z)$, and $\Sigma_u^u(t)$. $\Sigma_d^d$
contains similar terms along with $\Sigma_d^d(b)$ and $\Sigma_d^d(\tau)$
while $\Sigma_d^d(t)=0$~\cite{bbhmmt}. There are also contributions from $D$-term contributions to 
first/second generation squarks and sleptons which nearly cancel amongst themselves 
(due to sum of weak isospins/hypercharges equaling zero).
Once we are in parameter space where $m_{H_u}^2(m_{SUSY})\sim \mu^2\sim m_Z^2$, then the radiative
corrections $\Sigma_u^u$ may give the largest contribution to $\Delta_{EW}$. 

The largest of the $\Sigma_u^u$ almost always come from top squarks, where we find
\bea
\Sigma_u^u(\tst_{1,2} )&=&\frac{3}{16\pi^2}F(m_{\tst_{1,2}}^2)\times\nonumber \\ 
& &
\left[ f_t^2-g_Z^2\mp \frac{f_t^2 A_t^2-8g_Z^2(\frac{1}{4}-\frac{2}{3}x_W)\Delta_t}{m_{\tst_2}^2-m_{\tst_1}^2}
\right]
\label{eq:Siguu}
\eea
where $\Delta_t=(m_{\tst_L}^2-m_{\tst_R}^2)/2+m_Z^2\cos 2\beta(\frac{1}{4}-\frac{2}{3}x_W)$, 
$g_Z^2=(g^2+g^{\prime 2})/8$, $x_W\equiv \sin^2\theta_W$ and $F(m^2)=m^2\left(\log (m^2/Q^2)-1\right)$. 
In Ref. \cite{ltr}, it is shown that for the case of the $\tst_1$ contribution, as
$|A_t|$ gets large there is a suppression of $\Sigma_u^u(\tst_1)$ due to
a cancellation between terms in the square brackets of
Eq.~(\ref{eq:Siguu}).  For the $\tst_2$ contribution, a large
splitting between $m_{\tst_2}$ and $m_{\tst_1}$ yields a large
cancellation within $F(m_{\tst_2}^2)$
$\left(\log(m_{\tst_2}^2/Q^2)\to\log (m_{\tst_2}/m_{\tst_1})\to 1
\right)$ for $Q^2=m_{\tst_1}m_{\tst_2}$, leading also to suppression.  So while large $|A_t|$ values
suppress both top squark contributions to $\Sigma_u^u$, at the same time
they also lift up the value of $m_h$, which is near maximal for large,
negative $A_t$.  Combining all effects, one sees
that the same mechanism responsible for boosting the value of $m_h$ into
accord with LHC measurements can also suppress the $\Sigma_u^u$ contributions to EWFT, leading to a model
with electroweak naturalness. 

\begin{figure}[tbp]
\includegraphics[height=.3\textheight]{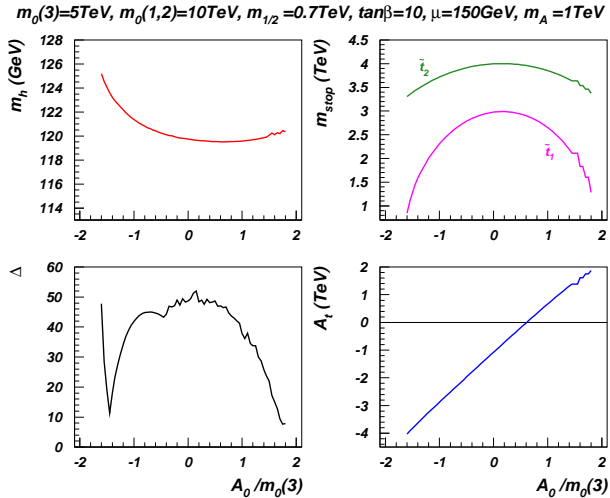}
\caption{Plot of {\it a}) $m_h$, {\it b}) $m_{\tst_{1,2}}$, {\it c}) $\Delta$ and {\it d}) $A_t$ versus
variation in $A_0$ for a model with $m_0(1,2)=10$ TeV, $m_0(3)=5$ TeV, $m_{1/2}=700$ GeV, $\tan\beta =10$
and $\mu =150$ GeV and $m_A=1$ TeV. 
\label{fig:1}}
\end{figure}

To illustrate these ideas, we adopt a simple benchmark point from the 2-parameter non-universal Higgs mass
SUSY model NUHM2\cite{nuhm2}, but with split generations, where $m_0(3)<m_0(1,2)$. 
In Fig. \ref{fig:1}, we take $m_0(3)=5$ TeV, $m_0(1,2)=10$ TeV, $m_{1/2}=700$ GeV, $\tan\beta =10$ with
$\mu=150$ GeV, $m_A=1000$ GeV and $m_t=173.2$ GeV. We allow the GUT scale parameter $A_0$ to vary, and calculate the 
sparticle mass spectrum using Isajet 7.83\cite{isajet}, which includes the new EWFT measure. 
In frame {\it a}), we plot the
value of $m_h$ versus $A_0$. While for $A_0\sim 0$ the value of $m_h\sim 120$ GeV, as $A_0$ moves towards $-2m_0(3)$, 
the top squark radiative contributions to $m_h$ increase, pushing its value up to 125 GeV.
(There is an expected theory error of $\pm 2$ GeV in our RGE-improved effective potential calculation of
$m_h$, which includes leading 2-loop effects\cite{hh}.)
At the same time, in frame {\it b}), we see the values of $m_{\tst_{1,2}}$ versus $A_0$. In this case, 
large values of $A_0$ suppress the soft terms $m_{Q_3}^2$ and $m_{U_3}^2$ via RGE running. But 
also large weak scale values of $A_t$ provide large mixing in the top squark mass matrix 
which suppresses $m_{\tst_1}$ and leads to an increased splitting between the two mass eigenstates 
which suppresses the top squark radiative corrections $\Sigma_u^u(\tst_2)$. 
The EWFT measure $\Delta\equiv\Delta_{EW}$ is shown in frame {\it c}), where we
see that while $\Delta\sim 50$ for $A_0=0$, when $A_0$ becomes large, then $\Delta$ drops to 10,
or $\Delta^{-1}= 10\%$ EWFT.
In frame {\it d}), we show the weak scale value of $A_t$ versus $A_0$ variation.
While the EWFT is quite low-- in the range expected for natural SUSY models-- we note that the top
squark masses remain above the TeV level, and in particular $m_{\tst_2}\sim 3.5$ TeV, in contrast to
previous natural SUSY expectations. 

\section{Sparticle spectrum}

The sparticle mass spectrum for this radiative NS benchmark point (RNS1)
is shown in Table \ref{tab:bm} for $A_0=-7300$ GeV. The heavier spectrum
of top and bottom squarks seem likely outside of any near-term LHC
reach, although in this case gluino\cite{bblt} and possibly heavy 
electroweak-ino\cite{wh} pair
production may be accessible to LHC14.  Dialing the $A_0$
parameter up to $-8$ TeV allows for $m_h=125.2$ GeV but increases EWFT
to $\Delta =29.5$, or 3.4\% fine-tuning. Alternatively, pushing $m_t$ up
to 174.4 GeV increases $m_h$ to $124.5$ GeV with 6.2\% fine-tuning;
increasing $\tan\beta$ to 20 increases $m_h$ to 124.6 GeV with 3.3\%
fine-tuning.  We show a second point RNS2 with $m_0(1,2)=m_0(3)=7.0$ TeV
and $\Delta =11.5$ with $m_h=125$ GeV; note the common sfermion mass parameter
at the high scale.  For comparison, we also show in
Table \ref{tab:bm} the NS2 benchmark from Ref. \cite{bbht}; in this case,
a more conventional light spectra of top squarks is generated leading to
$m_h=121.1$ GeV, but the model-- with $\Delta=23.7$-- has higher EWFT
than RNS1 or RNS2.

%
\begin{table}\centering
\begin{tabular}{lccc}
\hline
parameter & RNS1 & RNS2 & NS2 \\
\hline
$m_0(1,2)$      & 10000 & 7025.0 & 19542.2  \\
$m_0(3)$      & 5000 & 7025.0 & 2430.6  \\
$m_{1/2}$  & 700 & 568.3 & 1549.3  \\
$A_0$      & -7300 & -11426.6 & 873.2  \\
$\tan\beta$& 10 & 8.55 & 22.1  \\
$\mu$      & 150 & 150 & 150  \\
$m_A$      & 1000 & 1000 & 1652.7  \\
\hline
$m_{\tg}$   & 1859.0 & 1562.8 & 3696.8   \\
$m_{\tu_L}$ & 10050.9 & 7020.9 & 19736.2  \\
$m_{\tu_R}$ & 10141.6 & 7256.2 & 19762.6  \\
$m_{\te_R}$ & 9909.9 & 6755.4 & 19537.2  \\
$m_{\tst_1}$& 1415.9 & 1843.4 & 572.0  \\
$m_{\tst_2}$& 3424.8 & 4921.4 & 715.4  \\
$m_{\tb_1}$ & 3450.1 & 4962.6 & 497.3  \\
$m_{\tb_2}$ & 4823.6 & 6914.9 & 1723.8  \\
$m_{\ttau_1}$ & 4737.5 & 6679.4 & 2084.7  \\
$m_{\ttau_2}$ & 5020.7 & 7116.9 & 2189.1  \\
$m_{\tnu_{\tau}}$ & 5000.1 & 7128.3 & 2061.8  \\
$m_{\tw_2}$ & 621.3  & 513.9 & 1341.2  \\
$m_{\tw_1}$ & 154.2  & 152.7 & 156.1  \\
$m_{\tz_4}$ & 631.2 & 525.2 & 1340.4  \\ 
$m_{\tz_3}$ & 323.3 & 268.8 & 698.8   \\ 
$m_{\tz_2}$ & 158.5 & 159.2 & 156.2  \\ 
$m_{\tz_1}$ & 140.0 & 135.4 & 149.2  \\ 
$m_h$       & 123.7 & 125.0 & 121.1  \\ 
\hline
$\Omega_{\tz_1}^{std}h^2$ & 0.009 & 0.01 & 0.006  \\
$BF(b\to s\gamma)\times 10^4$ & $3.3$  & 3.3 & $3.6$  \\
$BF(B_s\to \mu^+\mu^-)\times 10^9$ & $3.8$  & 3.8 & $4.0$  \\
$\sigma^{SI}(\tz_1 p)$ (pb) & $1.1\times 10^{-8}$  & $1.7\times 10^{-8}$ & $1.8\times 10^{-9}$ \\
$\Delta$ & 9.7 & 11.5 & 23.7 \\
\hline
\end{tabular}
\caption{Input parameters and masses in~GeV units
for two {\it radiative natural SUSY} benchmark points and one NS point with $\mu =150$ GeV
and $m_t=173.2$ GeV.
}
\label{tab:bm}
\end{table}

The RNS model shares some features of generic NS models, but also includes
important differences. The several benchmark points shown in Table~\ref{tab:bm} 
imply that  RNS is characterized by:
\begin{itemize}
\item a higgsino mass $\mu\sim M_Z\sim 100-300$ GeV,
\item a light top squark $m_{\tst_1}\sim 1-2$ TeV,
\item a heavier top-squark $m_{\tst_2}\sim e m_{\tst_1}\sim 2-4$ TeV (here, $e\equiv 2.718...$),
\item $m_{\tg}\sim 1-5$ TeV,
\item first/second generation sfermions $\sim 5-20$ TeV.
\end{itemize}
While $\mu \sim m_Z$ as in usual NS models, the heavier top squarks and gluinos
implied by RNS allow for $m_h\simeq 125$ GeV but may make this model more difficult to detect at LHC than usual NS.

\begin{figure}[tbp]
\includegraphics[height=0.3\textheight]{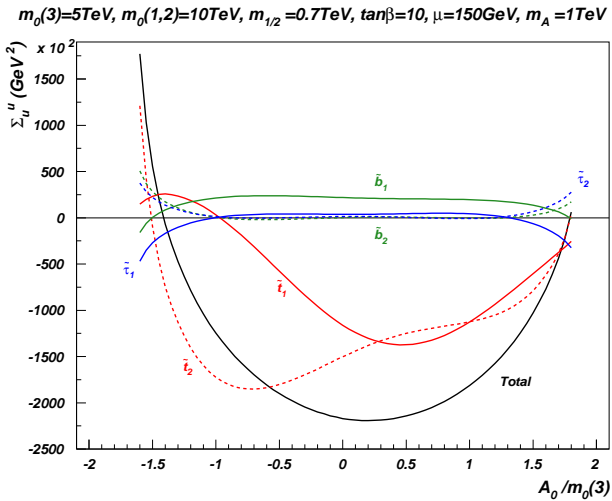}
\caption{Plot of third generation contributions to $\Sigma_u^u$ versus $A_0$ for benchmark point RNS1 
where solid curves come form the lighter mass eigenstate and dashed curves from the heavier. 
The black solid curve is $\Sigma_u^u$ which has summed over 
all contributions. 
\label{fig:2}}
\end{figure}

To illustrate how low EWFT comes about even with rather heavy top
squarks, we show in Fig. \ref{fig:2} the various third generation
contributions to $\Sigma_u^u$, where the lighter mass eigenstates are
shown as solid curves, while heavier eigenstates are dashed. The sum of
all contributions to $\Sigma_u^u$ is shown by the black curve marked
total. From the figure we see that for $A_0\sim 0$, indeed both top
squark contributions to $\Sigma_u^u$ are large and negative, leading to
a large value of $\Sigma_u^u(total)$, which will require large
fine-tuning in Eq.~(\ref{eq:loopmin}). As $A_0$ gets large negative, both
top squark contributions to $\Sigma_u^u$ are suppressed, and
$\Sigma_u^u(\tst_1 )$ even changes sign, leading to cancellations
amongst the various $\Sigma_u^u$ contributions.

\section{Radiative natural SUSY at colliders:} 

What chance does LHC have of detecting RNS? Unlike previous NS models, 
RNS has top and bottom squarks more typically in the $m_{\tst_1}\sim 1-2$ TeV
and $m_{\tst_2}\sim 2-4$ TeV range, likely beyond LHC reach. It also has 
light higgsinos $\tw_1^\pm$, $\tz_{1,2}$. 
While these latter particles can have substantial production cross sections at LHC, 
the mass gaps $m_{\tw_1}-m_{\tz_1}$ and $m_{\tz_2}-m_{\tz_1}$ are typically in the $10-50$ GeV
range. Thus, the visible decay products from $\tw_1^+\tw_1^-$ and $\tw_1^\pm\tz_2$ 
production tend to be at rather low energies, making observability difficult. The best bet 
may be if gluinos lie in the lower half of their expected range $m_{\tg}\sim 1-5$ TeV.
In this case, $\tg\to tb\tw_1$ or $t\bar{t}\tz_i$ decays occur, and one might expect
an observable rate for $\tg\tg\to 4t+\eslt$ signals. A portion of these events will contain
cascade decays to $\tz_2\to\tz_1\ell^+\ell^-$ and if the OS/SF dilepton pair can be reconstructed, 
then its distinctive invariant mass distribution bounded by $m_{\tz_2}-m_{\tz_1}$ may point to 
the presence of light higgsinos.

The hallmark of RNS and other NS models is the presence of light higgsinos with mass 
$m_{\tw_1},\ m_{\tz_{1,2}}\alt 200-300$ GeV. In this case, a linear $e^+e^-$ collider
operating with $\sqrt{s}\agt 2m_{\tw_1}$ would be a {\it higgsino factory} in addition to a Higgs
factory\cite{bbh,bl}! 
The soft decay products from $\tw_1\to\tz_1 f\bar{f}'$ decay are not problematic for
detection at an ILC, and will even be boosted as $\sqrt{s}$ increases well beyond threshold energy for creating
charginos pairs. The reaction $e^+e^-\to\tz_1\tz_2$ will also be distinctive.

\section{Mixed axion-higgsino cold dark matter:} 

In $R$-parity conserving SUSY models with higgsino-like WIMPs\cite{bbh}, the relic density is usually
a factor $10-15$ below the WMAP measured value of $\Omega_{CDM}h^2\simeq 0.11$. This is due to
a high rate of higgsino annihilation into $WW$ and $ZZ$ final states in the early universe.
Thus, the usual picture of thermally produced WIMP-only dark matter is inadequate for the case
of models with higgsino-like WIMPs. 

A variety of non-standard cosmological models have been proposed which can ameliorate this situation.
For instance, at least one relatively light modulus field is expected from string theory\cite{acharya}, 
and if the scalar field
decays after neutralino freeze-out with a substantial branching fraction into SUSY particles then 
it will typically augment the neutralino abundance\cite{mr}.

Alternatively, or in addition, if the strong $CP$ problem is solved by the Peccei-Quinn 
mechanism in a SUSY context, 
then we expect the presence of axions in addition to $R$-parity odd spin $\frac{1}{2}$ axinos $\ta$
and $R$-parity even spin-0 saxions $s$. In this case, dark matter could consist of two particles:
an axion-higgsino admixture\cite{ckls,blrs,bls}. 
The neutralinos are produced thermally as usual, but are also produced
via thermal production followed by cascade decays of axinos at high $T_R$. The late decay of
axinos into higgsinos can cause a re-annihilation of neutralinos at temperatures below freeze-out, substantially
augmenting the relic abundance. In addition, saxions can be produced both thermally at lower range of
PQ breaking parameter $f_a$, and via coherent oscillations at high $f_a$, and in fact may 
temporarily dominate the energy density of the universe. Their decay $s\to aa$ is expected to dominate
and would add to the measured $N_{eff}(\nu )$. Their decay $s\to \tg\tg$ or $\ta\ta$ would augment
the neutralino abundance, while late decays into SM particles such as $gg$ would dilute all relics
present at the time of decay. Exact dark matter abundances depend on the specific SUSY axion model 
and choices of PQMSSM parameters. It is possible one could have either axion or higgsino dominance of the
relic abundance, or even a comparable mixture. In the latter case, it may be possible to directly detect
both an axion and a higgsino-like WIMP.

\section{Conclusions:} 

Models of Natural SUSY are attractive in that they enjoy
low levels of EWFT, which arise from a low value of $\mu$ and possibly a
sub-TeV spectrum of top squarks and $\tb_1$.  In the context of the
MSSM, such light top squarks are difficult to reconcile with the LHC
Higgs boson discovery which requires $m_h\sim 125$ GeV. 
By imposing naturalness using $\Delta_{EW}$ with weak scale parameter inputs, we allow
for large cancellations in $m_{H_u}^2$ as it is driven to the weak scale.
In this case, for some range of $m_{H_u}^2(m_{GUT})>m_0$ (as in NUHM models), the weak
scale value of $m_{H_u}^2$ will be $\sim m_Z^2$ thus generating the
natural SUSY model radiatively. 
Models with a
large negative trilinear soft-breaking parameter $A_t$ can maximize the
value of $m_h$ into the $125$ GeV range without recourse to adding
exotic matter into the theory. The large value of $A_t$ also suppresses
1-loop top squark contributions to the scalar potential minimization condition
leading to models with low EWFT and a light Higgs scalar consistent with
LHC measurements. 
(More details on the allowable parameter space of RNS will be presented 
in Ref. \cite{bbhmmt}.)
The large negative $A_t$ parameter can arise from large negative $A_0$ at the GUT scale. 

While RNS may be difficult to detect at
LHC unless gluinos, third generation squarks or the heavier
electroweak-inos are fortuitously light, a linear $e^+e^-$ collider with
$\sqrt{s}\agt 2|\mu|$ would have enough energy to produce the hallmark
light higgsinos which are expected in this class of models. Since the model
predicts a lower abundance of higgsino-like WIMP dark matter in the standard cosmology, 
there is room for mixed axion-higgsino cold dark matter. 
The axions are necessary anyway if one solves the strong $CP$ problem via the PQ mechanism.

\begin{acknowledgments}
I thank my collaborators Vernon Barger, P. Huang, A. Lessa, D. Mickelson, A. Mustafayev,
S. Rajagopalan, W. Sreethawong and X. Tata.
I also thank Barbara Szczerbinska and Bhaskar Dutta for organizing an excellent CETUP workshop 
on dark matter physics.
HB would like to thank the Center for Theoretical Underground Physics
(CETUP) for hospitality while this work was completed.
This work was supported in part by the US Department of Energy, Office of High
Energy Physics.
\end{acknowledgments}


\end{document}